\documentclass[aps,pre,preprint,groupedaddress,12pt]{article}
\usepackage{geometry}
\usepackage{epsfig}
\usepackage{rotating}
\usepackage{color}
\usepackage{longtable}
\usepackage{multicol}
\usepackage[normalem]{ulem}
\usepackage{hyperref}
\usepackage{breakurl}
\usepackage{graphicx}
\usepackage{authblk}
\usepackage{breakcites}

\title{Analyzing time series activity of Twitter political spambots}

\author[1]{Oscar Fontanelli}
\author[2]{Aldo Venegas}
\author[3]{Ricardo Mansilla}
\affil[1]{Centro de Investigaciones Interdisciplinarias en Ciencias y Humanidades, Universidad Nacional Autónoma de México, oscarfontanelli@ciencias.unam.mx}

\affil[2]{Facultad de Ciencias, Universidad Nacional Autónoma de México}

\affil[3]{Centro de Investigaciones Interdisciplinarias en Ciencias y Humanidades and Centro Peninsular en Humanidades y Ciencias Sociales, UNAM}

\makeatletter
\renewcommand{\@maketitle}{%
{%
\thispagestyle{empty}%
\vskip-36pt%
{\raggedright\sffamily\bfseries\fontsize{20}{25}\selectfont \@title\par}%
\vskip10pt
{\raggedright\sffamily\fontsize{12}{16}\selectfont  \@author\par}
\vskip25pt%
}%
}%
\makeatother

\begin{document}

\flushbottom
\maketitle
%
%
\thispagestyle{empty}

The presence and complexity of political Twitter bots has increased in recent years, making it a very difficult task to recognize these accounts from real, human users. We intended to provide an answer to the following question: are temporal patterns of activity qualitatively different in fake and human accounts? We collected a large sample of tweets during the post-electoral conflict in the US in 2020 and performed supervised and non-supervised statistical learning techniques to quantify the predictive power of time-series features for human-bot recognition. Our results show that there are no substantial differences, suggesting that political bots are nowadays very capable of mimicking human behaviour. This finding reveals the need for novel, more sophisticated bot-detection techniques.

\section*{Introduction}
Digital social networks play an increasingly important role in political campaigns and conversations all around the world \cite{conover2011political, enli2013personalized, stier2018election}. Rapid diffusion of information and the possibility of direct interaction between politicians and citizens allow different political actors to spread their messages in a way that is efficient, speedy and at a very low cost \cite{alonso2016political, conway2015rise, dang2013investigation, vonderschmitt2012growing}. In recent years, the presence of fake or manipulated accounts designed to send spam has grown considerably within these social networks \cite{shao2018spread, shao2018anatomy}. In political contexts, these spambots are created and utilized to massively spread certain news (usually fake news), opinions, create artificial trends or support certain persons and organizations, thus contaminating organic conversation between normal users \cite{al2019fake, hegelich2016social,howard2016bots}. The effect they have is that of manipulating public opinion.

\hspace{1cm} Ever since the emergence of digital social networks there has been an interest - from public, private and academic points of view - in detecting and ultimately filtering out these kind of fake accounts \cite{beskow2018bot, chu2012detecting, cresci2015fame, davis2016botornot, davoudi2020towards, efthimion2018supervised, ferrara2016detection, rodriguez2020one, wei2019twitter}. Some of these techniques utilize time series features and they have been successful in telling human from bots with this information \cite{chavoshi2016identifying, chavoshi2017temporal, duh2018collective}, but as algorithms and tools for systematically detecting spambots have grown popular, so has increased the complexity and the capacity for mimicking human behaviour that these fake account possess. Nowadays, the problem of telling between normal and fake accounts requires very sophisticated techniques and it can be very difficult even with manual, human inspection. In this context we ask the following question: do human and fake accounts for political spamming still show qualitative differences in their time activity patterns? 

\hspace{1cm}In this work we analyze political conversations in the social network Twitter in the context of the post-electoral conflict in the US after the presidential elections of 2020. By means of supervised and non-supervised statistical learning techniques we quantify the predictive power of time series features for distinguishing bots from humans. Our results suggests that time activity features may be no longer useful for human-bot activity recognition in this social network.

\section*{Methodology}

\subsection*{Data Collection and Annotation}

\hspace{1cm}We utilized the public Twitter API to download 372,768 statuses (tweets, retweets, quotes or replies) containing either the hashtag $\#$StopTheSteal or $\#$MarchForTrump between November 5, 2020 and January 8, 2021.  These statuses were sent from 208,484 different accounts. Within this time interval we detected three periods of intense activity (early November, mid November and early January). Since we are interested in spambots and we want to compute certain properties of time series, we selected only accounts that posted five or more statuses within at least one of these three periods, thus remaining with 9,752 accounts.

\hspace{1cm}We have manually inspected the public profile of these accounts and classified them into two categories: normal accounts and suspect (of being spambots) accounts. These classification was done by three independent persons and was based solely on user metadata from the public profile, not taking into account any feature of their time series (since we don't want to construct ant kind of artificial correlation). This manual inspection yielded 8,782 normal accounts and 970 suspect accounts.

\subsection*{Time Series Characterization}

\hspace{1cm}For each account $k$ (assume it posted $n_k$ statuses) we extracted the series of time differences between consecutive statuses $\{\Delta t_{1}, \Delta t_{2},...,\Delta t_{(n-1)_{k}}\}$. We call this the \emph{time interval distribution} of account $k$. Next we computed for each account the first four moments of this distribution, $m_i = \sum (\Delta t_{j})^i/n_k$, for $i=1,2,3,4$.
We computed two different entropy measures for the time series: \emph{time entropy} \cite{ghosh2011entropy} and \emph{permutation entropy} \cite{bandt2002permutation}. Intuitively, an account that posts at perfectly regular time intervals (a possible indication of an automated account) will have an entropy of zero, while high values of the entropy indicate a more disordered time series. 



\begin{center}
\begin{table}[!t]
\begin{tabular}{lll}
\hline
Feature                 & Description                                             & Class        \\ \hline
n                       & Number of statuses with the desired hasthag   & temporal     \\
m1                      & First moment of time interval distribution              & temporal     \\
m2                      & Second moment of time interval distribution             & temporal     \\
m3                      & Third moment of time interval distribution              & temporal     \\
m4                      & Fourth moment of time interval distribution             & temporal     \\
time entropy            & Time entropy of time interval distribution              & temporal     \\
permutation entropy            & Permutation entropy of time interval distribution              & temporal     \\
followers count         & Number of followers                                     & non temporal \\
friends count           & Number of followed accounts                             & non temporal \\
statuses count          & Total number of statuses since the account was created  & non temporal \\
retweet ratio           & Fraction of statuses with the hashtag that are retweets & non temporal \\
friends-followers ratio & Number of friends / Number of followers                  & non temporal \\
age                     & Age of the account                                      & non temporal \\
screen name entropy     & Shannon entropy of the screen name                      & non temporal \\ \hline
\end{tabular}
\caption{Temporal and non-temporal features we utilized to train supervised classifiers.}
\label{tabla-1}
\end{table}
\end{center}

\vspace{-1cm}
\subsection*{Supervised classification}

\hspace{1cm}We selected two groups of features: those regarding time interval distribution (temporal features) and a list of characteristics that are typically used to identify Twitter bots (non temporal features) \cite{beskow2018bot, chu2012detecting,  cresci2015fame,  efthimion2018supervised, rodriguez2020one}. We show all the features we used on table \ref{tabla-1}. With these features we fitted three models for our binary classification (normal or suspect account): model 1, using only temporal features; model 2, using only non-temporal features; model 3, using all features. Our goal was to see if the predictive power and goodness of fit measures of these classification models significantly improve when we add temporal features. 
We fitted a random forest and a k-nearest neighbors classifier with models 1, 2 and 3. Hyperparameters of both classifiers were tuned by a grid search, choosing the best combination according to the area under the ROC curve criterion. The issue of the imbalanced data was approached with over and under sampling algorithms. 

\subsection*{Data Clustering}

\hspace{1cm}We performed k-means clustering on the Twitter users data with only temporal features. Our goal was to see whether suspect accounts tend to fall in the same cluster or group of clusters. If they do, this means that suspect accounts show similar temporal patterns of activity. We selected the value of $k$ (number of clusters) by looking for an inflection point in a plot of $k$ against total within sum of squares (the ``elbow method''). 

\begin{table}[!t]
\begin{tabular}{c|c|c|c|c|c|c|}
\cline{2-7}
\multicolumn{1}{l|}{}         & \multicolumn{3}{c|}{Random Forest}              & \multicolumn{3}{c|}{KNN}                        \\ \cline{2-7} 
\multicolumn{1}{l|}{}         & Precision & TP rate & Area under ROC & Precision & TP rate & Area under ROC \\ \hline
\multicolumn{1}{|c|}{model 1} & 0.813     & 0.127              & 0.605          & 0.775     & 0.218              & 0.576          \\
\multicolumn{1}{|c|}{model 2} & 0.715     & 0.614              & 0.744          & 0.738     & 0.544              & 0.703          \\
\multicolumn{1}{|c|}{model 3} & 0.730     & 0.614              & 0.758          & 0.758     & 0.519              & 0.740          \\ \hline
\end{tabular}
\caption{Metrics for models 1, 2 and 3 using random forest and KNN classifiers.}
\label{tabla-2}
\end{table}

\section*{Results and discussion}

\hspace{1cm}Table \ref{tabla-2} shows the main metrics for both classifiers and the three models. Notice how, because of the imbalance of the data, precision may not be a reliable metric. See for example results for random forest in model 1: precision seems to be high, but the extremely low value of the true positive (TP) rate suggests this is a highly-biased classifier towards the majority class (normal accounts). Considering this and the area under ROC curve, model 1 (only temporal features) does not seem to be a very good model. Even though the addition of temporal features increases area under ROC curve (from model 2 to model 3), we see that true positive stays the same or decreases; this means that model 2 (non temporal features) is actually better than model 3 for identifying spambots, even though model 3 is a slightly better model overall, at least for this particular dataset.

\begin{figure}[!h]
\centering
\includegraphics[width = 0.6\linewidth]{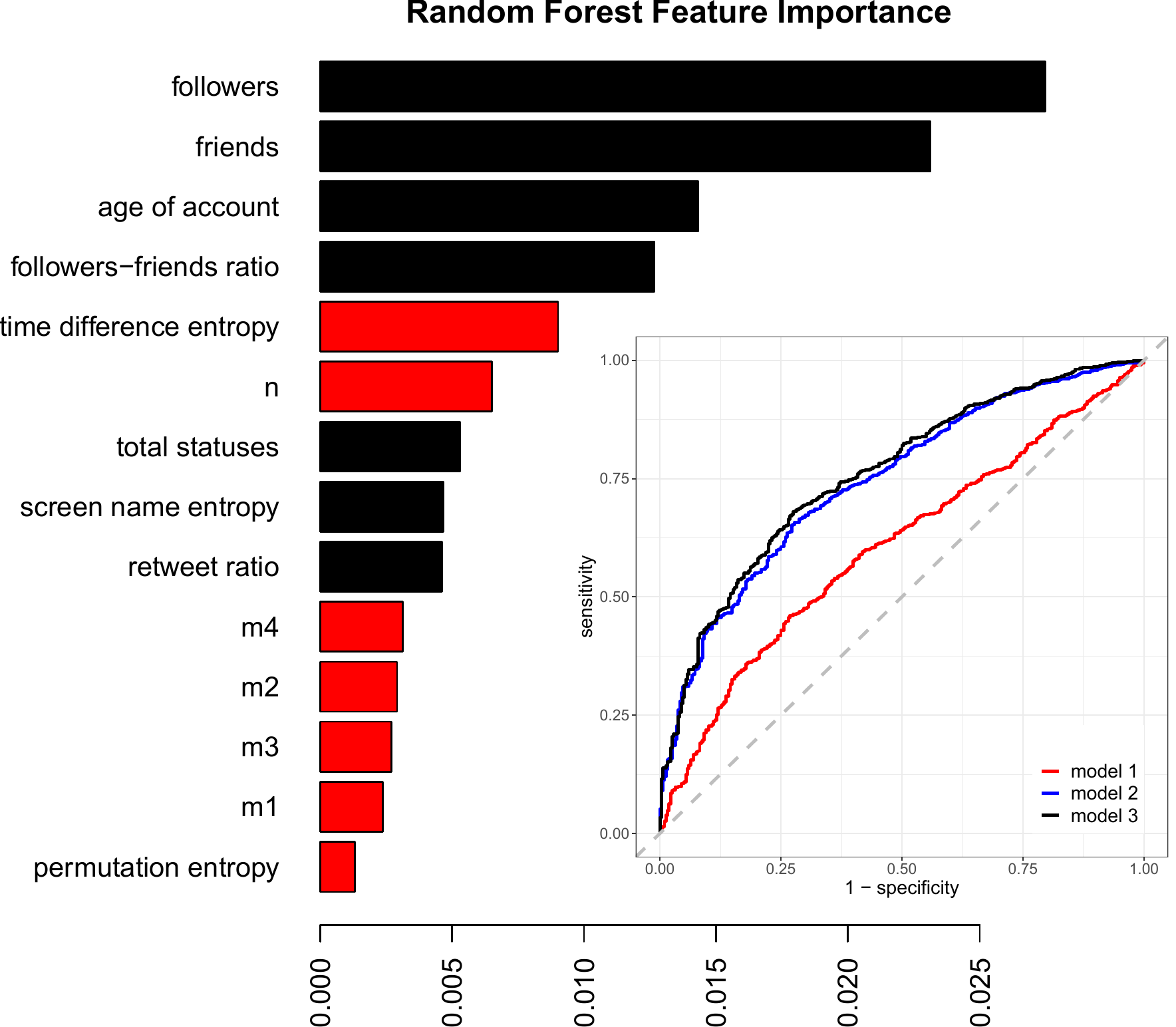}
\caption{Feature importance plot for model 3. Black bars correspond to non-temporal features, while red ones correspond to temporal features. We also show ROC curves for models 1, 2 and 3.}
\label{fig-importance}
\end{figure}

\hspace{1cm}Feature importance is a measure in the random forest classifier that measures the predictive power of each variable. We show in fig. \ref{fig-importance}  feature importance for model 3 (all features), along with ROC curves for all three models. Feature importance was computed using the mean decrease in Gini index. ROC curves for KNN classifier (not displayed here) show a qualitatively similar behavior. According to this metric, number of statuses and time entropy are the most important time-activity features, but all others features of this class are not very useful for classifying bots and humans.

\begin{figure}[!t]
\includegraphics[width = 0.9\linewidth]{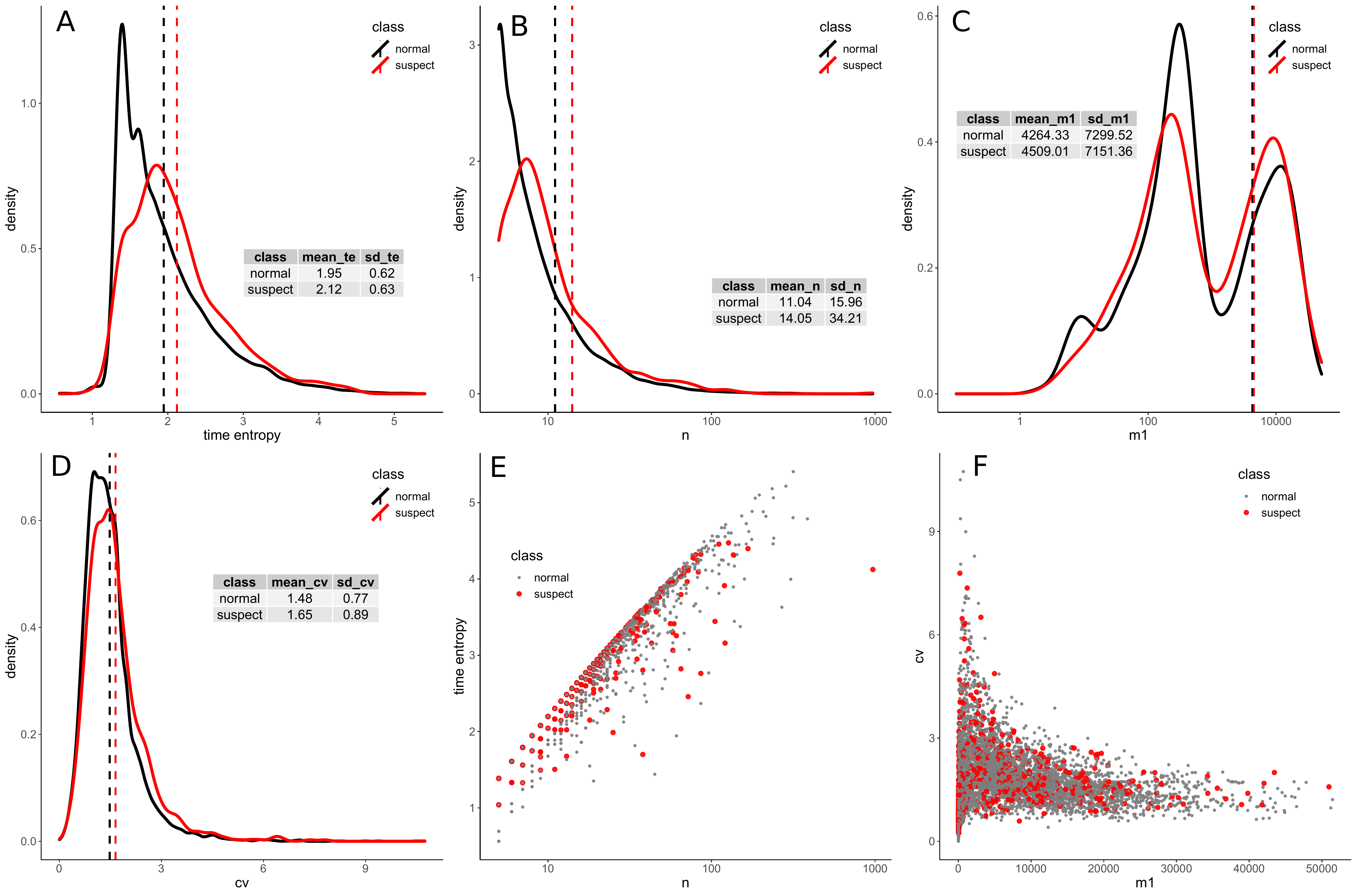}
\caption{A, B, C and D: density plots for time activity features in both classes; dotted lines indicate the mean for each group. E and F: scatter plots for time activity features. $n$ and $m1$ are in log scale.} 
\label{descriptive}
\end{figure}

\hspace{1cm}In order to visualize qualitative differences between the behaviors in both groups (normal and suspect accounts) we show in fig. \ref{descriptive} density and scatter plots for some of the temporal features. First we select the two temporal features with higher importance: time entropy (panel A) and number of statuses (panel B). While modes are visible different in both cases, distributions of both groups seem to be considerably overlapped. This overlapping is more evident for the distribution of the first moment (mean of time interval distribution, panel C) and for the distribution of the coefficient of variation (defined as $\sqrt{m2-m1^2}/m1$). On panel E we show a scatter plot of n vs time entropy; at first sight, there is no clear distinction between both groups. The same thing happens with a scatter plot $m1$ and coefficient of variation; accounts with highest normalized variability are normal accounts (gray dots), but the two groups overlap all other this phase space.

\begin{figure}[!t]
\includegraphics[width = 0.9\linewidth]{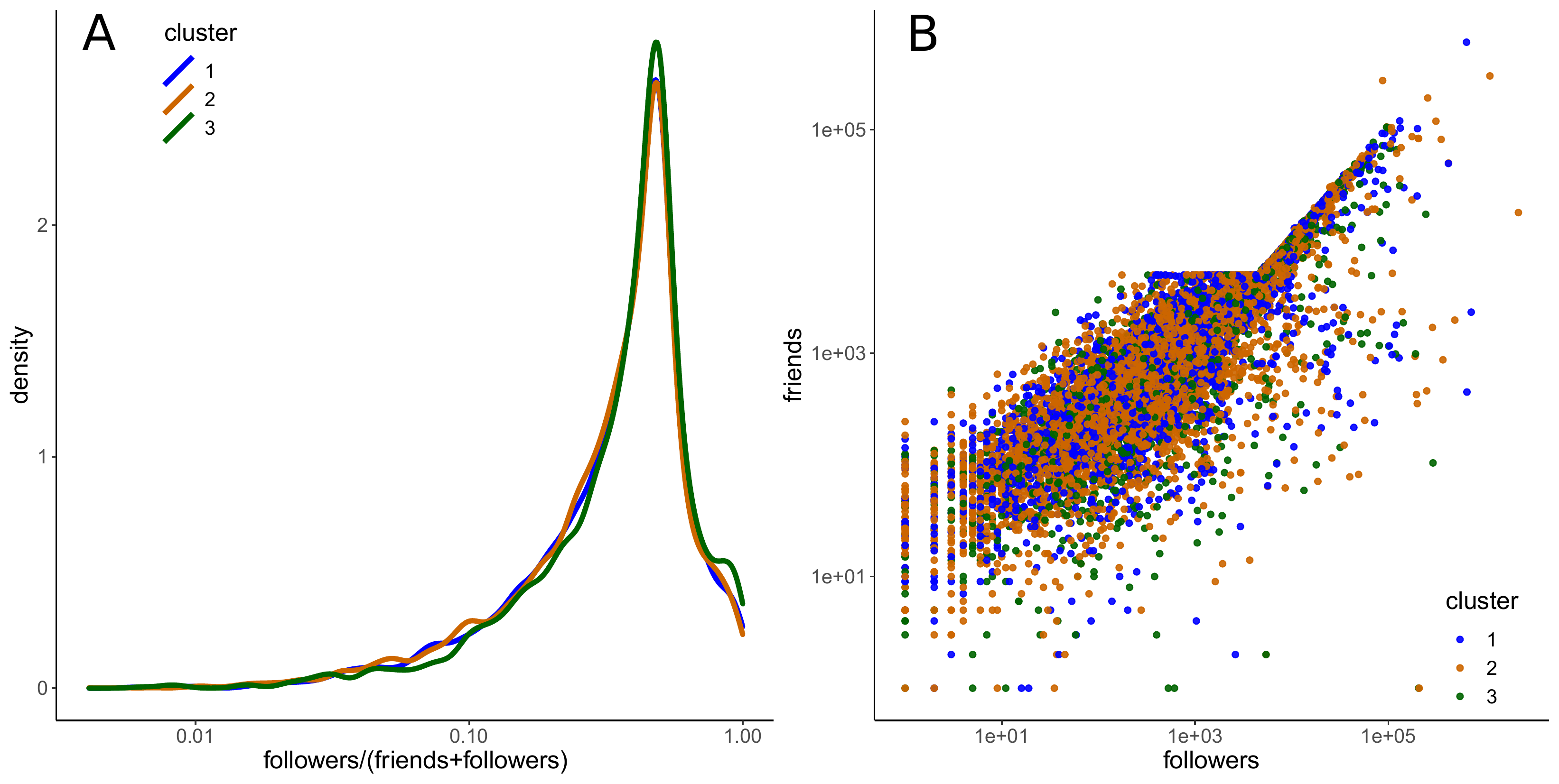}
\label{importance}
\caption{A: density plot for followers ratio for the three groups we found by K-mean clustering. B: scatter plot for followers vs friends (the two most important features according to the random forest classifier.} 
\label{knn}
\end{figure}

\hspace{1cm}For our unsupervised analysis, we chose $k=3$ for a k-nearest neighbors clustering according to the elbow criterion. When we look at non-temporal features (density for the fraction of followers and followers vs friends scatter plot in fig. \ref{knn} A and B) we see that all three groups mix together, indicating a lack of correlation between temporal and non-temporal features.

\begin{figure}[!h]
\includegraphics[width = 0.9\linewidth]{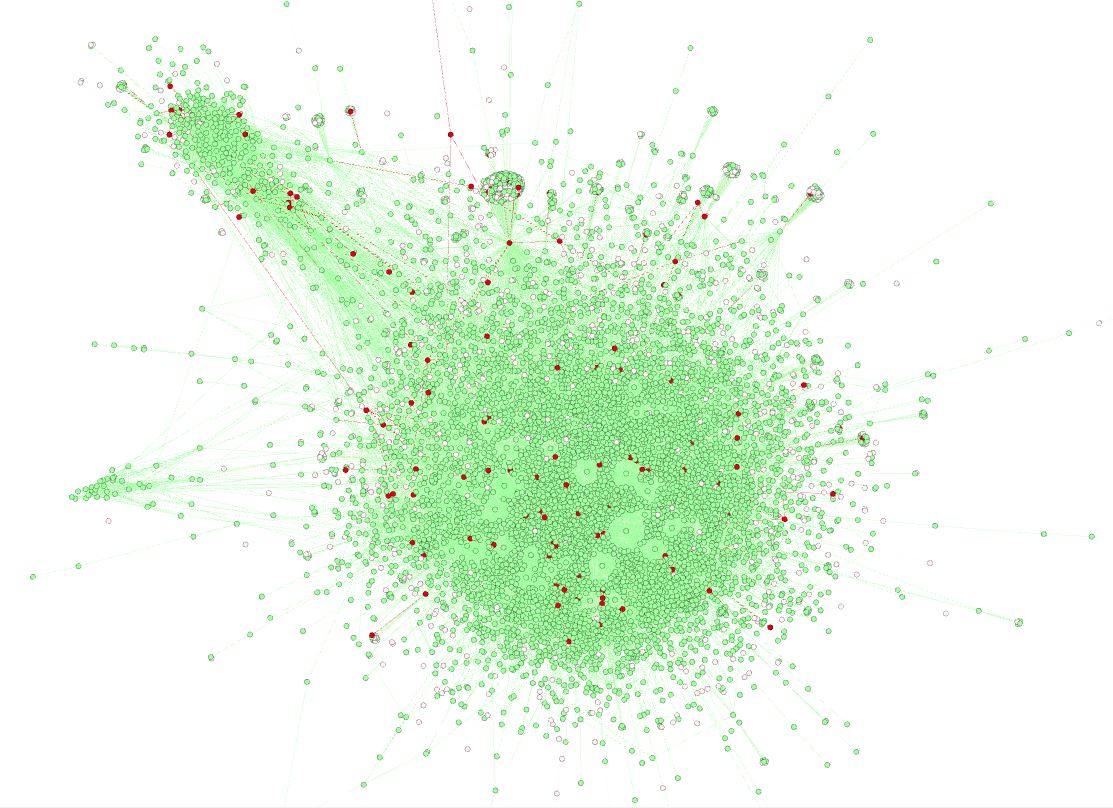}
\label{importance}
\caption{Retweet graph for $\#$StopTheSteal between November 5 and November 9, 2020. Nodes are Twitter accounts and they are connected if there is a retweet relationship between them. Green nodes are normal accounts and red nodes are suspect accounts. White nodes are accounts that are no longer available for analysis.} 
\label{network}
\end{figure}

\hspace{1cm}As an additional analysis, we constructed a retweet graph for the activity of $\#$StopTheSteal between November 5 and November 9, 2020. We selected accounts that tweeted four or more times in this period and established that two accounts are connected if one was retweeted by the other. This gives us a directed network where we can visualize structure within these digital communities. We show in fig. \ref{network} this network, with normal accounts in green and suspect accounts in red. Again, we see that both groups are completely mixed, indicating that suspect accounts are present all over the network. 

\section*{Conclusions}

\hspace{1cm}All results above discussed suggest that political spambots mimic normal, human temporal-behavior very well, thus putting into the question the validity of using time features to detect political spambots. This is a novel finding, since these kind of time activity features used to be useful for detecting bots on digital social networks. However, these fake accounts are constantly evolving, showing behaviours that are every time more similar to that of organic, human accounts. These results indicate the need to develop more sophisticated algorithms, techniques and tools to automatically detect and filter out these malicious account that pollute organic conversation on digital social networks and can potentially manipulate public opinion.  

\section*{Acknowledgments}

\hspace{1cm}OF was a grant holder of the Dirección General de Asuntos del Personal Académico, Universidad Nacional Autónoma de México - Postdoctoral Scholarships Program at CEIICH, UNAM.

\vspace{1cm}

\bibliographystyle{abbrv}
\bibliography{references}

\end{document}